\documentclass[12pt]{article}
\usepackage{amsmath,hyperref,graphicx,slashed,bbold}

\newcommand{\be}{\begin{equation}}
\newcommand{\ee}{\end{equation}}
\newcommand{\bea}{\begin{eqnarray}}
\newcommand{\eea}{\end{eqnarray}}
\newcommand{\beas}{\begin{eqnarray*}}
\newcommand{\eeas}{\end{eqnarray*}}

\begin{document}
\begin{titlepage}

\medskip

\begin{center}

{\Large Induced Lorentz Violation on a Moving Braneworld}

\vspace{12mm}

\renewcommand\thefootnote{\mbox{$\fnsymbol{footnote}$}}
Daniel Kabat${}^{1,3}$\footnote{daniel.kabat@lehman.cuny.edu},
Marcelo Nomura${}^{2,3}$\footnote{mail@marcelonomura.com}

\vspace{6mm}

${}^1${\small \sl Department of Physics and Astronomy} \\
{\small \sl Lehman College, City University of New York} \\
{\small \sl 250 Bedford Park Blvd.\ W, Bronx, NY 10468, USA}

\vspace{3mm}

${}^2${\small \sl Physics Department} \\
{\small \sl City College of the City University of New York} \\
{\small \sl 160 Convent Avenue, New York, NY 10031, USA}

\vspace{3mm}

${}^3${\small \sl Graduate School and University Center, City University of New York} \\
{\small \sl  365 Fifth Avenue, New York, NY 10016, USA}

\end{center}

\vspace{12mm}

\noindent
We consider a braneworld scenario in which a flat 4-D brane, embedded in $M^{3,1} \times S^1$, is moving on or spiraling around the $S^1$.
Although the induced metric on the brane is 4-D Minkowski, the would-be Lorentz symmetry of the brane is broken globally by the compactification.
As recently pointed out this means causal bulk signals can propagate superluminally and even backwards in time according to brane observers.
Here we consider the effective action on the brane induced by loops of bulk fields.  We consider a variety of self-energy and vertex corrections due
to bulk scalars and gravitons and show that bulk loops with non-zero winding generate UV-finite Lorentz-violating terms in
the 4-D effective action.  The results can be accommodated by the Standard Model Extension, a general framework for Lorentz-violating effective field theory.

\end{titlepage}

\setcounter{footnote}{0}
\renewcommand\thefootnote{\mbox{\arabic{footnote}}}

\hrule
\tableofcontents
\bigskip
\hrule
\addtolength{\parskip}{8pt}
\section{Introduction}
The simplest braneworld scenario posits a spacetime of the form $M^{3,1} \times S^1$, with a single extra dimension compactified on a circle of radius $R$.  The brane, assumed to be at a fixed position
on the $S^1$, has a Minkowski metric induced on its worldvolume.  In this scenario worldvolume Lorentz invariance is an exact symmetry, inherited from a symmetry of the underlying 5-D spacetime.

A straightforward generalization of this scenario allows the brane to either move on or spiral around the $S^1$.  This generalization might seem quite innocuous.  The induced metric on the brane is still
4-D Minkowski, so it would seem that brane observers might be hard-pressed to find any evidence that their brane has been boosted or rotated into the compact direction.

From a different perspective, however, the effects of this generalization are quite dramatic.  Compactification on $S^1$ preserves an $SO(3,1)$ symmetry that acts on the directions orthogonal to the $S^1$.
Once the brane is moving on or spiraling around the $S^1$ this exact $SO(3,1)$ symmetry no longer aligns with the would-be Lorentz symmetry of the brane worldvolume.  Although there is no local indication of the violation,
worldvolume Lorentz invariance is broken globally by the compactification.  Without Lorentz invariance all bets are off, and indeed \cite{Greene:2022urm,Greene:2022uyf,Polychronakos:2022uam} showed that bulk signals can
propagate faster than light and even backwards in time according to brane observers.  Fortunately causality -- a more robust feature -- remains intact, inherited from the causality of the underlying
5-D spacetime.

Here we consider the effects of virtual bulk particles in this generalized braneworld scenario.  Such particles can leave a moving or rotated brane, propagate around the compact dimension, and return.  Bulk loops have no reason to respect
worldvolume Lorentz symmetry and might be expected to induce Lorentz-violating terms in the brane effective action.  We will see that this is indeed the case.  We focus on Lorentz-violating dimension-4 terms in the effective action,
especially the electron self-energy and the electron -- photon vertex, and show that bulk loops induce specific Lorentz-violating terms with finite, calculable coefficients.  These terms are part of the Standard
Model Extension, a general framework for Lorentz-violating effective field theories developed in \cite{Colladay:1996iz,Colladay:1998fq}.  There are stringent experimental bounds on
the Lorentz-violating coefficients which have been tabulated in \cite{Kostelecky:2008ts}.

An outline of this paper is as follows.  In section \ref{sect:compactification} we describe the braneworld scenario we will consider.  Section \ref{sect:propagator} discusses the
propagator for a bulk scalar field.  In sections \ref{sect:electron-scalar} and \ref{sect:vertex} we evaluate corrections to the electron self-energy and the electron -- photon vertex
due to a bulk scalar loop.  Section \ref{sect:scalar} considers the self-energy for a scalar field on the brane induced by a bulk scalar.  Sections \ref{sect:electron-graviton} and \ref{sect:scalar-graviton} evaluate the electron and scalar self-energy induced by a bulk graviton loop.  We conclude in section \ref{sect:conclusions} with a discussion of experimental bounds and future directions.

Cosmological implications of this scenario have been studied in \cite{Bernardo:2023eda} and a different approach to braneworld Lorentz violation has been developed in \cite{Dai:2023zsx}.

\section{Boosted and rotated branes\label{sect:compactification}}
Consider a 5-D spacetime ${\cal M}^{3,1} \times S^1$.  To describe this we begin from 5-D Minkowski space ${\cal M}^{4,1}$ with coordinates
and metric
\bea
&& X^{M} = (X^\mu,Z) \qquad M = 0,\ldots,4 \qquad \mu = 0,\ldots,3 \\[3pt]
\nonumber
&& \eta_{MN} = {\rm diag}(+ - \cdots -)
\eea
We obtain an $S^1$ by periodically identifying the $Z$ coordinate, $Z \approx Z + 2 \pi R$.  It's convenient to describe this identification as
\be
\label{compactify}
X^M \approx X^M + A^M \qquad A^M = (0,0,0,0,2 \pi R)
\ee
These upper-case coordinates define the preferred frame for the compactification, with an exact $SO(3,1)$ symmetry that acts on the coordinates $X^\mu$.

The standard braneworld scenario would be to place a 4-D braneworld at rest at $Z = 0$.  We are instead interested in braneworld which is moving
in the $Z$ direction and / or has been rotated into the $Z$ direction.  To describe this we transform to a new frame with lower-case coordinates $x^M$ via
\be
x^M = L^M{}_N X^N
\ee
Here $L^M{}_N$ is an $SO(4,1)$ transformation that acts non-trivially on the $Z$ coordinate.  In the $x^M$ coordinates there is a boosted and / or
rotated identification
\be
\label{identify}
x^M \approx x^M + a^M \qquad\quad a^M = L^M{}_N A^N
\ee
We set
\be
x^M = (x^\mu,z)
\ee
and imagine a braneworld at $z = 0$.  The coordinates $x^M$ can be thought of as co-moving and / or co-rotated with the brane.
Since all we've done is a 5-D Lorentz transformation, in the co-moving coordinates the metric still has the form
\be
ds^2 = \eta_{\mu\nu} dx^\mu dx^\nu - dz^2
\ee
The compactification is hidden in the identification (\ref{identify}).  Thus the induced metric on the brane is 4-D Minkowski, however the $SO(3,1)$
symmetry of the brane metric does not align with the $SO(3,1)$ symmetry that is preserved by the compactification (\ref{compactify}).
Instead worldvolume Lorentz symmetry is broken globally by the compactification, which leads to the curious possibilities of superluminal
and even backwards-in-time signaling explored in \cite{Greene:2022urm,Greene:2022uyf,Polychronakos:2022uam}.

From the brane point of view it's natural to decompose $a^M$ into components tangent and normal to the brane, so we set
\be
a^M = (a^\mu,2 \pi r)
\ee
$a^\mu$ becomes a preferred 4-vector on the brane, which shows that 4-D Lorentz symmetry on the brane is spontaneously broken.  The fifth
component $2 \pi r$ is a scalar on the brane.  In the calculations below we will find it useful to work with the combination
\be
b^\mu = {a^\mu \over 2 \pi r}
\ee
Up to Lorentz transformations on the brane there are three cases to consider.
\begin{enumerate}
\item {\bf Timelike $b^\mu$}

In this case we can go to a reference frame on the brane in which $b^\mu$ only has a time component.  This can be obtained directly from
(\ref{compactify}) by boosting with velocity $\beta$ in the $Z$ direction.
\be
\left(\begin{array}{c} T \\ Z \end{array}\right) = \left(\begin{array}{cc} \gamma & \gamma \beta \\ \gamma \beta & \gamma \end{array}\right)
\left(\begin{array}{c} t \\ z \end{array}\right)
\ee
This leads to
\be
a^\mu = (-\gamma \beta \, 2 \pi R, 0, 0, 0) \qquad r = \gamma R
\ee
Note that
\be
b^\mu = (-\beta, 0, 0, 0)
\ee
with
\be
b^2 = \beta^2 \in (0,1)
\ee
or alternatively
\be
1 - b^2 = {1 \over \gamma^2} \in (0,1)
\ee
This corresponds to the ``boost-like isotropic'' case discussed in \cite{Polychronakos:2022uam}.  As seen on the brane, bulk signals propagate isotropically in all directions
at superluminal speeds.

\item {\bf Spacelike $b^\mu$}

In this case we can go to a reference frame on the brane in which $b^\mu$ only has an $x$ component.  This can be obtained directly from
(\ref{compactify}) by rotating through an angle $\theta$ in the $XZ$ plane.
\be
\left(\begin{array}{c} X \\ Z \end{array}\right) = \left(\begin{array}{cc} \cos \theta & -\sin \theta \\ \sin \theta & \cos \theta \end{array}\right)
\left(\begin{array}{c} x \\ z \end{array}\right)
\ee
This leads to
\be
a^\mu = (0, \sin \theta \, 2 \pi R, 0, 0) \qquad r = \cos \theta R
\ee
Note that
\be
b^\mu = (0, \tan \theta, 0, 0)
\ee
with
\be
b^2 = - \tan^2 \theta \in (-\infty,0)
\ee
or alternatively
\be
1 - b^2 = {1 \over \cos^2 \theta} \in (1,\infty)
\ee
This corresponds to the ``tilt-like anisotropic'' case discussed in \cite{Polychronakos:2022uam}.  As seen on the brane, bulk signals propagate superluminally
in the $x$ direction and at the speed of light in perpendicular directions.

\item {\bf Lightlike $b^\mu$}

Finally we consider the case of lightlike $b^\mu$.\footnote{We are grateful to Alexios Polychronakos and Massimo Porrati for bringing up this possibility.}  This can be obtained starting from (\ref{compactify}) by making a Lorentz
rotation in the $X^- \, Z$ plane.\footnote{The case of a rotation in the $X^+ \, Z$ plane proceeds similarly.}  Here we've introduced light-front coordinates $X^\pm = T \pm X$ with
\be
\label{LFmetric}
ds^2 = dX^+ dX^- - \vert d{\bf Y} \vert^2 - dZ^2
\ee
The form of the Lorentz transformation is a little unfamiliar.  Introducing a parameter $\lambda \in {\mathbb R}$ it takes the form
\be
\label{J-Z}
\left(\begin{array}{c} T \\ X \\ Z \end{array}\right) = \left(\begin{array}{ccc} 1 + {1 \over 2} \lambda^2 & {1 \over 2} \lambda^2 & \,\lambda \\
- {1 \over 2} \lambda^2 & 1 - {1 \over 2} \lambda^2 & - \lambda \\
\lambda & \lambda & \,1 \end{array}\right)
\left(\begin{array}{c} t \\ x \\ z \end{array}\right)
\ee
or equivalently
\be
\label{J-Z2}
\left(\begin{array}{c} X^+ \\ X^- \\ \!\!Z \end{array}\right) = \left(\begin{array}{ccc} 1 & 0 & 0 \\
\lambda^2 & 1 & 2 \lambda \\
\lambda & 0 & 1 \end{array}\right)
\left(\begin{array}{c} x^+ \\ x^- \\ \!\!z \end{array}\right)
\ee
To see that this is the appropriate Lorentz transformation note that it leaves $X^+$ invariant, $X^+ = x^+$, so it acts on $X^- \, Z$ planes.  Also it preserves the metric (\ref{LFmetric}),
with $ds^2 = dx^+ dx^- - \vert d{\bf y} \vert^2 - dz^2$.  Applying the (inverse of) the transformation (\ref{J-Z}) gives
\be
a^\mu = (-\lambda 2 \pi R, \lambda 2 \pi R, 0, 0) \qquad\quad r = R
\ee
So the radius is unchanged, while
\be
b^\mu = (-\lambda,\lambda,0,0)
\ee
is indeed a null vector on the brane.

A null vector has no invariant length, so one can go to an infinitely-boosted frame in which $b^\mu = 0$.  This restores conventional Lorentz invariance on the brane.  However if any matter (e.g.\ CMB photons) is present on the brane one may not wish to perform an infinite boost.  In section \ref{sect:propagator} we show that when $\lambda$ is non-zero bulk signals can have a
negative light-front velocity in the $x^-$ direction.  With respect to Minkowski time this means that as seen on the brane a bulk signal can travel
faster than light and even backwards in time in the $x$ direction.  For further discussion of the geometry of this case see appendix \ref{appendix:null}.
\end{enumerate}
Note that in all three cases we have $b^2 < 1$.  The range $-\infty < b^2 < 0$
is tilt-like, $b^2 = 0$ is null, and $0 < b^2 < 1$ is boost-like.  Alternatively we can say that we always have $1 - b^2 > 0$.\footnote{This is important for convergence
of the loop integrals we encounter below.}  The range $0 < 1 - b^2 < 1$ is boost-like, $1 - b^2 = 1$ is null,
and $1 < 1 - b^2 < \infty$ is tilt-like.

\section{Bulk scalar propagator\label{sect:propagator}}
We expect that bulk loops should induce Lorentz-violating terms on the brane.  Before turning to explicit calculations we start with a
discussion of the bulk propagator.  We focus on bulk scalar fields for simplicity.

The retarded propagator for a bulk field was discussed in \cite{Greene:2022urm} while the static Green's function was studied in \cite{Greene:2011fm}.
Here we consider the Feynman propagator.  It's straightforward to impose
the appropriate periodicity $(x^\mu,z) \approx (x^\mu + a^\mu, z + 2 \pi r)$ using a winding sum (equivalently, a sum over image charges).  In position space
this gives the propagator for a bulk scalar of mass $\mu$ as
\be
\Delta = \sum_{w = -\infty}^\infty \int {d^4k \over (2\pi)^4} \int {dq \over 2\pi} \, {i \over k^2 - q^2 - \mu^2 + i\epsilon}
e^{-i k \cdot (x - w a)} e^{i q (z - 2 \pi r w)}
\ee
It's convenient to shift
\be
q \rightarrow q + {k \cdot a \over 2 \pi r} = q + k \cdot b
\ee
so that
\be
\Delta = \sum_{w = -\infty}^\infty \int {d^4k \over (2\pi)^4} \int {dq \over 2\pi} \, {i \over k^2 - (q + k \cdot b)^2 - \mu^2 + i\epsilon}
e^{-i k \cdot x} e^{i (q + k \cdot b) z} e^{-i q 2 \pi r w}
\ee
We set $z = 0$ since we will only be interested in bulk propagation that starts and ends on the brane.  Also we work in momentum
space along the brane, which amounts to dropping $\int {d^4k \over (2\pi)^4} e^{-i k \cdot x}$.  Then we are left with the winding-sum form
for the bulk propagator,
\be
\Delta = \sum_{w = -\infty}^\infty \int {dq \over 2\pi} \, {i \over k^2 - (q + k \cdot b)^2 - \mu^2 + i\epsilon} \, e^{-i q 2 \pi r w}
\ee
We can switch from a sum over windings to a sum over Kaluza-Klein momentum using the Poisson resummation identity
\be
\label{Poisson}
\sum_w \int {dq \over 2\pi} f(q) e^{-i q 2 \pi r w} = {1 \over 2 \pi r} \sum_{n = -\infty}^\infty f\Big({n \over r}\Big)
\ee
This puts the bulk propagator in the form
\be
\label{MomentumSum}
\Delta = {1 \over 2 \pi r} \sum_{n = -\infty}^\infty {i \over k^2 - (k \cdot b + {n \over r})^2 - \mu^2 + i \epsilon}
\ee
It's clear that $b^\mu \not= 0$ breaks 4-D Lorentz invariance.  We can look for poles in the propagator and read off the dispersion
relation for the Kaluza-Klein tower, to see how it is modified from the perspective of a moving or rotated brane \cite{Greene:2011fm,Polychronakos:2022uam}.  There are three cases to consider.

\begin{enumerate}
\item  {\bf Timelike $b^\mu$}

In this case we set $b^\mu = (-\beta, 0, 0, 0)$ and $k^\mu = (\omega,{\bf k})$.  The propagator has poles at 
\be
\omega = \gamma\left[{\beta \gamma n \over r} \pm \sqrt{{\gamma^2 n^2 \over r^2} + \vert {\bf k} \vert^2 + \mu^2}\,\right] \mp i \epsilon \qquad
n \in {\mathbb Z}
\ee
One branch of solutions has $\omega > 0$ and a pole that is displaced slightly below the real axis.  The other branch has $\omega < 0$
and a pole that is displaced slightly above the real axis.  Although we don't have 4-D Lorentz invariance, the poles are displaced in the standard
way that allows for a Wick rotation to Euclidean signature.  One can check that there are no tachyons from a 4-D perspective, $\omega^2 - \vert {\bf k}\vert^2 \geq 0$.  Finally we can evaluate the group velocity
\be
v_g = {d\omega \over d k} = {\gamma \vert {\bf k} \vert \over \sqrt{\big({\gamma n \over r}\big)^2 + \vert {\bf k} \vert^2 + \mu^2}}
\ee
This makes it clear that wave propagation is isotropic, with a velocity $0 \leq v_g < \gamma$ that exceeds the speed of light if $\vert {\bf k} \vert$
is sufficiently large.

\item {\bf Spacelike $b^\mu$}

In this case we set $b^\mu = (0, \tan \theta, 0, 0)$ and $k^\mu = (\omega,k_x,{\bf k}_\perp)$.  The propagator has poles at 
\be
\omega = \pm \sqrt{\Big({k_x \over \cos \theta} - {n \sin \theta \over r}\Big)^2 + \vert {\bf k}_\perp \vert^2 + \Big( {n \cos \theta \over r} \Big)^2 + \mu^2}
\mp i \epsilon \qquad n \in {\mathbb Z}
\ee
Again one branch of solutions has $\omega > 0$ and a pole that is displaced slightly below the real axis, while the other branch has $\omega < 0$
and a pole that is displaced slightly above the real axis, so we can perform Wick rotation in the standard way.  One can check that there are no tachyons from a 4-D perspective, $\omega^2 - k_x^2 - \vert {\bf k}_\perp\vert^2 \geq 0$.
Finally the group velocity is anisotropic.  For a wave propagating in the $x$ direction
\be
v_{gx} = \left.{d \omega \over d k_x}\right\vert_{{\bf k}_\perp = 0} = {\left({k_x \over \cos \theta} - {n \sin \theta \over r}\right) \over \cos \theta \sqrt{\Big({k_x \over \cos \theta} - {n \sin \theta \over r}\Big)^2 + \Big( {n \cos \theta \over r} \Big)^2 + \mu^2}}\\
\ee
while for a wave propagating in one of the perpendicular directions
\be
v_{g\perp} = \left.{d \omega \over d k_\perp}\right\vert_{k_x = 0} = {\vert k_\perp \vert \over \sqrt{\vert k_\perp \vert^2 + \left({n \over r}\right)^2 + \mu^2}}
\ee
In the perpendicular directions we have the familiar group velocity for a Kaluza-Klein tower of particles with masses $\mu_n^2 = \left({n \over r}\right)^2 + \mu^2$.  In the $x$ direction we have a group velocity
$0 \leq \vert v_{gx} \vert < {1 \over \cos \theta}$ that exceeds the speed of light if $\vert k_x \vert$ is sufficiently large.

\item {\bf Lightlike $b^\mu$}

For the lightlike case we set $b^\mu = (-\lambda,\lambda,0,0)$.  It's convenient to introduce light-front coordinates on the brane.
\be
x^\pm = t \pm x \qquad k^\pm = \omega \pm k_x
\ee
We'll interpret $\tau = x^+$ as light-front time and the conjugate momentum $k_+ = {1 \over 2} k^-$ as light-front energy.  The propagator has poles at
\be
k^- = {1 \over k^+} \left[\big(\lambda k^+ - {n \over  r}\big)^2 + \vert {\bf k}_\perp \vert^2 + \mu^2\right]
\ee
This fixes the dispersion relation.
As usual there are two branches of solutions.  Positive frequency modes have $k^+ > 0$ and $k^- > 0$, while negative frequency modes have $k^+ < 0$
and $k^- < 0$.  Given a positive-frequency plane-wave solution
\be
e^{-i\big({1 \over 2} k^- x^+ + {1 \over 2} k^+ x^- - {\bf k}_\perp \cdot {\bf x}_\perp\big)}
\ee
a stationary-phase approximation lets us read off the group velocities with respect to light-front time.
\bea
\label{vminus}
&& v^- = {dx^- \over d\tau} = - {\partial k^- \over \partial k^+} = {\vert {\bf k}_\perp \vert^2 + \big({n \over r}\big)^2 + \mu^2 \over \big(k^+\big)^2} - \lambda^2 \\
\label{vperp}
&& {\bf v}_\perp = {d{\bf x}_\perp \over d\tau} = {1 \over 2} {\partial k^- \over \partial {\bf k}_\perp} = {{\bf k}_\perp \over k^+}
\eea
In the transverse directions we have conventional light-front kinematics.\footnote{Although the transverse kinematics look conventional, with respect to Minkowski time
the limiting transverse velocity is $\sqrt{1 + \lambda^2}$.  See appendix \ref{appendix:null}.}  In the longitudinal direction there is a shift which allows the longitudinal velocity to be negative,
$-\lambda^2 < v^- < \infty$.  This means that in Minkowski coordinates bulk signals can travel faster than light and even backwards in time in the $x$ direction.  To see this
note that in Minkowski coordinates a trajectory $x^- = - \lambda^2 x^+$ corresponds to
\be
\label{plusx}
x = {1 + \lambda^2 \over 1 - \lambda^2} \, t
\ee
The Minkowski velocity is superluminal for $0 < \lambda^2 < 1$.  The velocity diverges at $\lambda^2 = 1$ and becomes negative for $\lambda^2 > 1$, which as in
\cite{Greene:2022uyf} indicates that the signal is traveling backwards in time.  For further discussion of this case see appendix \ref{appendix:null}.

\end{enumerate}

\section{Electron self-energy\label{sect:electron-scalar}}
The world-volume metric induced on the brane is 4-D Minkowski, even if the brane is boosted or rotated in the $Z$ direction.  Particles
that solely propagate on the brane are not sensitive to the breaking of 4-D Lorentz invariance and it would be reasonable to describe these
``standard model'' particles using an effective action with 4-D Lorentz symmetry.  However particles that propagate in the bulk can leave the
brane, travel around the compactification manifold, and return.  Such particles notice the global breaking of 4-D Lorentz invariance by the compactification
and loops of such particles should induce Lorentz-violating terms in the 4-D effective action.

\begin{figure}
\begin{center}
\includegraphics[height=3cm]{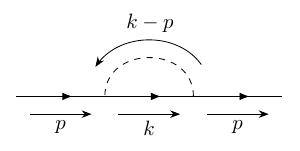}
\end{center}
\caption{One-loop electron self-energy arising from a Yukawa coupling to a bulk scalar.\label{diagram}}
\end{figure}

Here we study this effect, beginning with the simple example of radiative corrections to the electron self-energy.  We imagine a real bulk scalar field
$\chi$ of mass $\mu$ that has a Yukawa coupling to the electron.  We describe the coupled system with the action
\be
S = \int d^5x \, \left[{1 \over 2} \partial_M \chi \partial^M \chi - {1 \over 2} \mu^2 \chi^2\right]
+ \int d^4x \, \left[ \bar{\psi} \left(i \gamma^\mu \partial_\mu - m\right) \psi - \lambda \bar{\psi} \psi \chi\vert_{z = 0} \right]
\ee
Note that the coupling $\lambda$ has units $({\rm mass})^{-1/2}$.  The diagram we wish to consider is shown in Fig.\ \ref{diagram}.

Our goal is to evaluate the diagram and expand in powers of the external momentum $p$.  In this way we will make contact with the Standard Model Extension, a general effective field theory framework for
Lorentz-violating effects developed in \cite{Colladay:1996iz,Colladay:1998fq}.  The basic diagram is easy to write down.  Writing the brane-to-brane bulk propagator with a sum over Kaluza-Klein momentum
as in (\ref{MomentumSum}) we have
\be
\label{ElectronSigma}
i\Sigma = {\lambda^2 \over 2 \pi r} \sum_{n=-\infty}^\infty \int {d^4k \over (2\pi)^4} \, {\slashed{k} + m \over k^2 - m^2 + i\epsilon} \,
{1 \over (k-p)^2 - \big((k - p) \cdot b + {n \over r}\big)^2 - \mu^2 + i \epsilon}
\ee
As pointed out in section \ref{sect:propagator}, even though the bulk propagator is not Lorentz
invariant, it still has poles that allow for a standard Wick rotation.  So we Wick rotate in the usual way, setting
\be
\label{Wick}
k_E = (-i k^0,{\bf k}) \qquad d^4k = i d^4k_E \qquad k^2 = -k_E^2
\ee
with a similar rotation for all other 4-vectors.  We introduce a pair of Schwinger parameters $s$, $t$ to represent the propagators
via the identity
\be
\label{Schwinger}
\int_0^\infty ds \, e^{-As} = {1 \over A}
\ee
It's convenient to use a frame in which only the first component of $b_E$ is non-zero.
\bea
\nonumber
&& b_E = (b_E,{\bf 0}) \\
\nonumber
&& p_E = (p_E,{\bf p}_E) \\
\label{momenta}
&& k_E = (k_E,{\bf k}_E) \\
\nonumber
&& \gamma_E = (\gamma_E,\boldsymbol{\gamma}_E)
\eea
This leads to
\bea
\nonumber
i \Sigma & = & {i \lambda^2 \over 2 \pi r} \sum_{n = -\infty}^\infty \int_0^\infty ds \int_0^\infty dt \int {d^4k_E \over (2\pi)^4}
\left(-\gamma_E k_E - \boldsymbol{\gamma}_E \cdot {\bf k}_E + m\right)
\exp\left[-s\left(k_E^2 + \vert {\bf k}_E \vert^2 + m^2\right)\right] \\
\nonumber
& & \exp\left\lbrace-t\left[(1 + b_E^2) (k_E - p_E)^2 - {2 n \over r} b_E(k_E - p_E) + \vert {\bf k}_E - {\bf p}_E \vert^2 + \left({n \over r}\right)^2 + \mu^2\right]\right\rbrace \\
\eea
The momentum integrals are Gaussian and lead to the rather tedious expression
\bea
\nonumber
i \Sigma & = & {i \lambda^2 \over 32 \pi^3 r} \sum_{n = -\infty}^\infty \int_0^\infty ds \int_0^\infty dt
{1 \over (s + t)^{3/2} \big(s + t (1 + b_E^2)\big)^{1/2}} \\
\nonumber& & \left(-{t(1 + b_E^2) \over s + t(1 + b_E^2)} \gamma_E p_E - {t \over s + t} \boldsymbol{\gamma}_E \cdot {\bf p}_E -
{t \over s + t(1 + b_E^2)} \gamma_E b_E {n \over r} + m\right) \\
\nonumber
& & \exp\bigg\lbrace -sm^2 - t\Big(\left({n \over r}\right)^2 + \mu^2\Big) - {s t \over s + t(1+b_E^2)}\left(p_E^2(1 + b_E^2) + 2 b_E p_E {n \over r}\right) \\
\label{iSigma}
& & \qquad - {s t \over s + t} \vert {\bf p}_E \vert^2 + {t^2 \over s + t(1 + b_E^2)} b_E^2 \left({n \over r}\right)^2\bigg\rbrace
\eea

Now we expand in powers of the external momentum $p$.  At zeroth order, after continuing back to Lorentzian signature and restoring Lorentz covariance, we find
\bea
\nonumber
i \Sigma^{(0)} & = & {i \lambda^2 \over 32 \pi^3 r} \sum_{n = -\infty}^\infty \int_0^\infty ds \int_0^\infty dt
{1 \over (s + t)^{3/2} \big(s + t (1 - b^2)\big)^{1/2}} \left({t \over s + t(1 - b^2)} {n \over r} \, \slashed{b} + m\right) \\
& & \exp\bigg\lbrace -sm^2 - t\Big(\left({n \over r}\right)^2 + \mu^2\Big) - {t^2 b^2 \over s + t(1 - b^2)} \left({n \over r}\right)^2\bigg\rbrace
\eea
The term proportional to $m$ is Lorentz invariant and therefore not interesting to us.  The term proportional to $\slashed{b}$ has the potential to violate Lorentz invariance,
but it vanishes once the sum over $n$ is performed.  This follows from a symmetry: the underlying expression (\ref{iSigma}) is invariant under
$b_E \rightarrow -b_E$, $n \rightarrow -n$ which implies that
only even powers of $b^\mu$ can appear.

At first order in $p^\mu$, after continuing back to Lorentzian signature and restoring Lorentz covariance, we find
\bea
\nonumber
i \Sigma^{(1)} & = & {i \lambda^2 \over 32 \pi^{5/2}} \sum_{w = -\infty}^\infty \int_0^\infty ds \int_0^\infty dt
\left({t^{1/2} \over (s + t)^3} \slashed{p} - {2 (\pi r w)^2 s \over t^{1/2} (s+t)^4} b_\mu b_\nu \gamma^\mu p^\nu\right) \\
\label{iSigma1}
& & \exp\left\lbrace-sm^2 -t\mu^2 - {s + t(1-b^2) \over t(s+t)} (\pi r w)^2\right\rbrace
\eea
Here we've used the identity (\ref{Poisson}) in reverse to replace the momentum sum with a winding sum and an integral over $q$.  The integral over
$q$ is Gaussian and leads to (\ref{iSigma1}).

Working with a winding sum is advantageous for the following reasons.
\begin{itemize}
\item
Lorentz symmetry is broken globally by the compactification.  Particle trajectories with $w = 0$ are not sensitive to the breaking and are guaranteed to
respect Lorentz invariance.  Indeed in (\ref{iSigma1}) we see that the term with $w = 0$ is proportional to $\slashed{p}$.
\item
Ultraviolet divergences can only arise from the $w = 0$ term in the sum, since non-zero winding means the loop can never shrink to a point.  Indeed in
(\ref{iSigma1}) we see that for $w \not= 0$ the exponential in the second line serves to cut off the short-distance regime $s,\,t \rightarrow 0$.
\end{itemize}
Since we are only interested in Lorentz-violating terms, we could simply discard the $w = 0$ term to obtain a finite result.  However we might as well
discard all terms proportional to $\slashed{p}$.  This means discarding the first term in parenthesis in (\ref{iSigma1}) as well as the trace part of $b_\mu b_\nu$.
In this way we obtain the Lorentz-violating contribution
\bea
\nonumber
i \Sigma^{(1)}_{LV} & = & -{i \lambda^2 r^2 \over 16 \pi^{1/2}} \left(b_\mu b_\nu - {1 \over 4} \eta_{\mu\nu}b^2\right) \gamma^\mu p^\nu
\sum_{w = -\infty}^\infty \int_0^\infty ds \int_0^\infty dt \,
{s w^2 \over t^{1/2} (s + t)^4} \\
\label{iSigma1LV}
& & \quad \exp\left\lbrace-sm^2 -t\mu^2 - {s + t(1-b^2) \over t(s+t)} (\pi r w)^2\right\rbrace
\eea

This corresponds to a Lorentz-violating term in the 4-D effective Lagrangian for the electron.  In the notation of \cite{Colladay:1998fq} the relevant
term is ${\cal L} = i c_{\mu\nu} \bar{\psi} \gamma^\mu \partial^\nu \psi$ which makes a contribution $i c_{\mu\nu} \gamma^\mu p^\nu$
to $i \Sigma$.  Comparing to (\ref{iSigma1LV}) we can read off the Lorentz violating coefficient $c_{\mu\nu}$, which can be conveniently presented as
\be
\label{c}
c_{\mu\nu} = - {1 \over 16 \pi^2} \, {\lambda^2 \over \pi r} \left(b_\mu b_\nu - {1 \over 4} \eta_{\mu\nu}b^2\right) I_1
\ee
where we've defined\footnote{We rescaled the Schwinger parameters, $s \rightarrow \pi^2 r^2 s$ and $t \rightarrow \pi^2 r^2 t$.}
\be
\label{In}
I_n = {1 \over \sqrt{\pi}} \sum_{w = -\infty}^\infty \int_0^\infty ds \int_0^\infty dt \,
{s^n w^2 \over t^{1/2} (s + t)^4} \exp\left\lbrace-s(\pi m r)^2 -t(\pi \mu r)^2 - {s + t(1-b^2) \over t(s+t)} w^2\right\rbrace
\ee
The induced coefficients $c_{\mu\nu}$ are real, dimensionless, traceless and symmetric.  They make a Lorentz-violating but CPT-even contribution
to the effective action.

\begin{figure}
\begin{center}
\includegraphics[height=9cm]{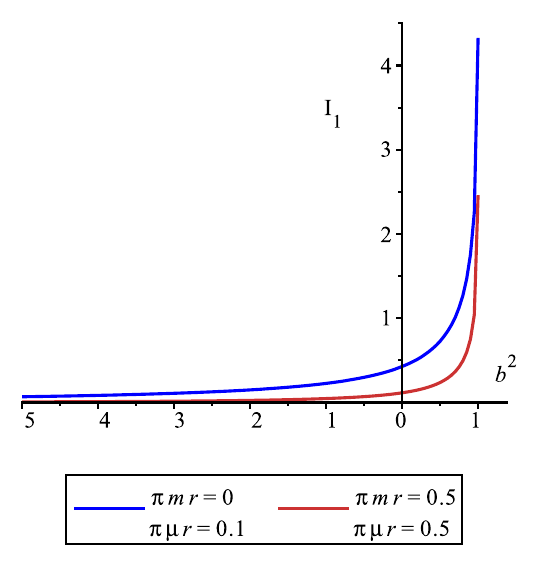}
\end{center}
\caption{The quantity $I_1$ appearing in the electron self-energy as a function of $b^2$.\label{fig:I1}}
\end{figure}

We can think of (\ref{c}) as a product of a loop factor ${1 \over 16 \pi^2}$, a dimensionless coupling ${\lambda^2 \over \pi r}$, a tensor structure
$b_\mu b_\nu - {1 \over 4} \eta_{\mu\nu} b^2$, and a function $I_1$ of the dimensionless parameters
$b^2$, $\pi m r$, $\pi \mu r$.  As can be seen in Fig.\ \ref{fig:I1}, $I_1$ is an increasing function of $b^2$.  It vanishes as $b^2 \rightarrow - \infty$ and
(perhaps despite appearances) approaches a finite limit as $b^2 \rightarrow 1$.

The expression for $I_1$ simplifies if we set $m = 0$ (a massless fermion on the brane) and $b^2 = 0$
(a small boost and / or rotation).  Then the sum and integrals can be performed and the behavior for small and large $\mu r$ can be extracted.\footnote{For a massless fermion on the brane we must keep the
bulk scalar mass $\mu$ non-zero to avoid an IR divergence.}  This leads to
\be
b^2 \approx 0,\, m \approx 0 \, : \quad I_1 \approx \left\lbrace\begin{array}{ll}
\displaystyle {1 \over 6} \log {1 \over \mu r} & \quad \hbox{\rm as $\mu r \rightarrow 0$} \\[15pt]
\displaystyle {\pi \over 3} \, \mu r e^{-2 \pi \mu r} & \quad \hbox{\rm as $\mu r \rightarrow \infty$}
\end{array}\right.
\ee

\section{Electron -- photon vertex\label{sect:vertex}}
\begin{figure}
\begin{center}
\includegraphics{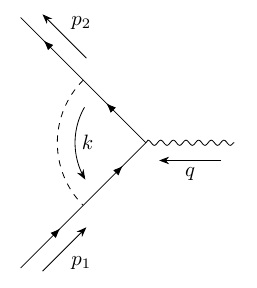}
\end{center}
\caption{One-loop vertex correction due to a bulk scalar.\label{vertex_correction}}
\end{figure}

Next we consider the one-loop correction to the electron -- photon vertex due to a bulk scalar.  The diagram is shown in Fig.\ \ref{vertex_correction}.

Suppressing the external polarizations and writing the bulk propagator with a momentum sum as in (\ref{MomentumSum}), the diagram is
\be
\label{vertex_diagram}
i \Gamma^\mu = -{e \lambda^2 \over 2 \pi r} \sum_{n = -\infty}^\infty \int {d^4k \over (2\pi)^4} \, {\slashed{p}_2 + \slashed{k} + m \over (p_2 + k)^2 - m^2 + i \epsilon} \, \gamma^\mu \,
{\slashed{p}_1 + \slashed{k} + m \over (p_1 + k)^2 - m^2 + i \epsilon} \, {1 \over k^2 - (k \cdot b + {n \over r})^2 - \mu^2 + i \epsilon}
\ee
We can evaluate this similarly to the electron self-energy.
We Wick rotate, introduce a series of Schwinger parameters
\bea
\nonumber
&& \frac{1}{(p_{1 E} + k_{E})^{2} + m^{2}}
=
\int_{0}^{\infty} ds_{1} \, e^{- s_{1}\left[(p_{1 E} + k_{E})^{2} + m^{2}\right]} \\
&& \frac{1}{(p_{2 E} + k_{E})^{2} + m^{2}}
=
\int_{0}^{\infty} ds_{2} \, e^{- s_{2}\left[(p_{2 E} + k_{E})^{2} + m^{2}\right]} \\
\nonumber
&& \frac{1}{k_{E}^{2} + \left(k_{E} \cdot b_{E} - \frac{n}{r}\right)^{2} + \mu^{2}}
=
\int_{0}^{\infty}dt \, e^{- t\left[k_{E}^{2} + \left(k_{E} \cdot b_{E} - \frac{n}{r}\right)^{2} + \mu^{2}\right]},
\eea
and evaluate the Gaussian integral over $k_E^\mu$.  This gives
\bea
\nonumber
&&i \Gamma^\mu={i e \lambda^2 \over 2 \pi r} \sum_{n = -\infty}^\infty \int_0^\infty ds_1 \int_0^\infty ds_2 \int_0^\infty dt \,
{1 \over 16 \pi^2} \sqrt{\det h} \, e^{-s_1 (p_{1E}^2 + m^2) - s_2 (p_{2E}^2 + m^2) -t ({n^2 \over r^2} + \mu^2)} \\
\nonumber
&& \qquad e^{h_{\mu\nu}v^\mu v^\nu} \left[{1 \over 2} h_{\alpha\beta} \gamma_E^\alpha \gamma_E^\mu \gamma_E^\beta
+ \left(p_{2E} \cdot \gamma_E - h_{\alpha\beta}v^\alpha \gamma_E^\beta - m\right) \gamma_E^\mu \left(p_{1E} \cdot \gamma_E - h_{\gamma\delta}v^\gamma\gamma_E^\delta - m\right)\right] \\
\label{iGamma}
\eea
where we've introduced the convenient notation
\bea
\nonumber
&& h_{\mu\nu} = {\rm diag.} \left({1 \over s_1 + s_2 + t(1 + b_E^2)}, {1 \over s_1 + s_2 + t}, {1 \over s_1 + s_2 + t}, {1 \over s_1 + s_2 + t}\right) \\[5pt]
&& v^\mu = s_1 p_{1E}^\mu + s_2 p_{2E}^\mu - t b_E^\mu {n \over r}
\eea
Now we expand in powers of the external momenta.  At leading (zeroth) order, after continuing back to Lorentzian signature,
switching to a winding sum for the bulk propagator and doing a bit of Dirac algebra, we find
\bea
\nonumber
&& i\Gamma^{(0)\mu} = i e \lambda^2 \sum_{w = -\infty}^\infty \int_0^\infty ds_1 \int_0^\infty ds_2 \int_0^\infty dt {1 \over (4 \pi t)^{1/2} \big(4\pi(s_1 + s_2 + t)\big)^2}
e^{- s_1 m^2 - s_2 m^2 - t \mu^2} \\
\nonumber
&& \qquad \exp \left\lbrace -{s_1 + s_2 + t(1-b^2) \over t(s_1 + s_2 + t)} \pi^2 r^2 w^2 \right\rbrace \left[{1 \over s_1 + s_2 +t} \gamma^\mu + m^2 \gamma^\mu - {(\pi r w)^2 \over (s_1 + s_2 + t)^2}
\big(2 \slashed{b} b^\mu - b^2 \gamma^\mu\big)\right] \\
\eea
We drop all Lorentz-invariant terms, which includes the UV-divergent terms with $w = 0$.  Setting $s = s_1 + s_2$ we're left with the Lorentz-violating contribution
\bea
\nonumber
i\Gamma^{(0)\mu}_{LV} & = & -{i e \lambda^2 r^2 \over 16 \pi^{1/2}} \big(b_\alpha b_\beta - {1 \over 4} \eta_{\alpha\beta} b^2\big) \gamma^\alpha \eta^{\beta\mu}
\sum_{w = -\infty}^\infty \int_0^\infty ds \int_0^\infty dt {s w^2 \over t^{1/2} (s+t)^4} \\
&& \exp \left\lbrace - s m^2 - t \mu^2 - {s + t(1-b^2) \over t(s + t)} \pi^2 r^2 w^2 \right\rbrace
\eea
This pairs nicely with (\ref{iSigma1LV}) to produce a gauge-invariant but Lorentz-violating dimension-4 term in the effective action, namely
\be
{\cal L} = i c_{\mu\nu} \bar{\psi} \gamma^\mu D^\nu \psi \qquad D_\mu = \partial_\mu - i e A_\mu
\ee
The coefficient $c_{\mu\nu}$ is given in (\ref{c}).  Since we stopped at zeroth order in the momentum this outcome, required by gauge invariance and Ward identities, can be thought of as a consistency check on our results.
Expanding (\ref{iGamma}) beyond zeroth order in the external momenta would give higher-derivative corrections to the
electron -- photon vertex.

\section{Scalar self-energy\label{sect:scalar}}
\begin{figure}
\begin{center}
\includegraphics[height=3cm]{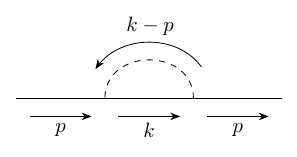}
\end{center}
\caption{One-loop scalar self-energy arising from a cubic coupling to a bulk scalar.\label{diagram2}}
\end{figure}

Having calculated the one-loop Lorentz-violating correction to the self-energy of an electron, we now perform a similar calculation for a real scalar field $\phi$
on the brane with a cubic coupling to a bulk scalar $\chi$.  We start from the action
\be
S = \int d^5x \, \left[{1 \over 2} \partial_M \chi \partial^M \chi - {1 \over 2} \mu^2 \chi^2\right]
+ \int d^4x \, \left[ {1 \over 2} \partial_\mu \phi \partial^\mu \phi - {1 \over 2} m^2 \phi^2 - {1 \over 2} g \phi^2 \chi\vert_{z = 0} \right]
\ee
Note that the coupling $g$ has units $({\rm mass})^{+1/2}$.  The diagram we wish to consider is shown in Fig.\ \ref{diagram2}.

Writing the brane-to-brane bulk propagator with a sum over Kaluza-Klein momentum as in (\ref{MomentumSum}), the diagram is
\be
{g^2 \over 2 \pi r} \sum_{n=-\infty}^\infty \int {d^4k \over (2\pi)^4} \, {1 \over k^2 - m^2 + i \epsilon}
{1 \over (k-p)^2 - \big((k - p) \cdot b + {n \over r}\big)^2 - \mu^2 + i \epsilon}
\ee
We Wick rotate to Euclidean signature as in (\ref{Wick}) and introduce a pair of Schwinger parameters as in (\ref{Schwinger}).  Parametrizing the
Euclidean momenta as in (\ref{momenta}) and performing the Gaussian integral over $k_E$ we find
\bea
\nonumber
&& {i g^2 \over 32 \pi^3 r} \sum_{n = -\infty}^\infty \int_0^\infty ds \int_0^\infty dt
{1 \over (s + t)^{3/2} \big(s + t (1 + b_E^2)\big)^{1/2}} \\
\nonumber
& & \exp\bigg\lbrace -sm^2 - t\Big(\left({n \over r}\right)^2 + \mu^2\Big) - {s t \over s + t(1+b_E^2)}\left(p_E^2(1 + b_E^2) + 2 b_E p_E {n \over r}\right) \\
& & \qquad - {s t \over s + t} \vert {\bf p}_E \vert^2 + {t^2 \over s + t(1 + b_E^2)} b_E^2 \left({n \over r}\right)^2\bigg\rbrace
\eea

Now we expand in powers of the external momentum $p$.  At zeroth order the result is Lorentz invariant and can be ignored.  At first order the sum over Kaluza-Klein momentum vanishes because it is odd under $n \rightarrow -n$.
At second order, after continuing back to Lorentzian signature and restoring Lorentz covariance, we find
\bea
\nonumber
&& {i g^2 \over 32 \pi^{5/2}} \sum_{w = -\infty}^\infty \int_0^\infty ds \int_0^\infty dt
\left({s t^{1/2} \over (s + t)^3} p^2 - {2 (\pi r w)^2 s^2 \over t^{1/2} (s+t)^4} b_\mu b_\nu p^\mu p^\nu\right) \\
\label{iSigmaScalar}
& & \exp\left\lbrace-sm^2 -t\mu^2 - {s + t(1-b^2) \over t(s+t)} (\pi r w)^2\right\rbrace
\eea
Again we've used the identity (\ref{Poisson}) in reverse to replace the momentum sum with a winding sum and an integral over $q$.  The integral over
$q$ is Gaussian and leads to (\ref{iSigmaScalar}).  The first term in parenthesis is Lorentz invariant and can be dropped.  The second term can be
matched to a Lorentz-violating term in the effective action \cite{Colladay:1998fq}
\be
{\cal L} = {1 \over 2} k_{\mu\nu} \partial^\mu \phi \partial^\nu \phi
\ee
with a traceless coefficient $k_{\mu\nu}$.
Removing the Lorentz-invariant trace from the second term in (\ref{iSigmaScalar}) we identify
\be
\label{k}
k_{\mu\nu} = - {1 \over 16 \pi^2} \, g^2 \pi r \left(b_\mu b_\nu - {1 \over 4} \eta_{\mu\nu}b^2\right) I_2
\ee
where $I_n$ is defined in (\ref{In}).  The induced coefficients $k_{\mu\nu}$ are real, dimensionless, traceless and symmetric.  They make a Lorentz-violating but CPT-even contribution
to the effective action.

\begin{figure}
\begin{center}
\includegraphics[height=9cm]{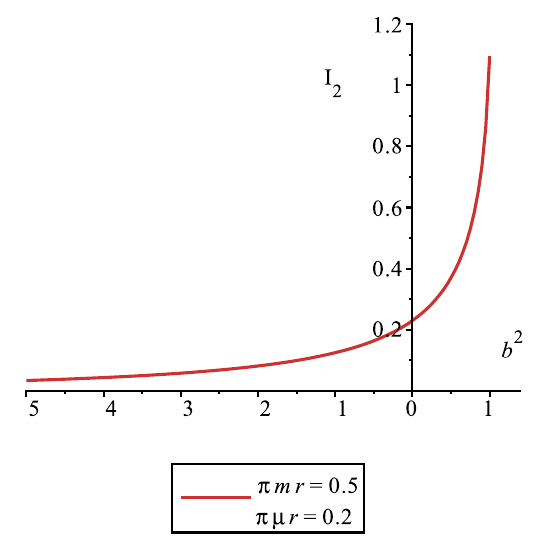}
\end{center}
\caption{The quantity $I_2$ appearing in the scalar self-energy as a function of $b^2$.\label{fig:I2}}
\end{figure}

We can think of (\ref{k}) as a product of a loop factor ${1 \over 16 \pi^2}$, a dimensionless coupling $g^2 \pi r$, a tensor structure $b_\mu b_\nu - {1 \over 4} \eta_{\mu\nu} b^2$, and a function $I_2$
of the dimensionless parameters $b^2$, $\pi m r$, $\pi \mu r$.  As can be seen in Fig.\ \ref{fig:I2}, $I_2$ is an increasing function of $b^2$.  It vanishes as $b^2 \rightarrow - \infty$ and
approaches a finite limit as $b^2 \rightarrow 1$.

The expression for $I_2$ simplifies if we set $m = 0$ (a massless scalar on the brane) and $b^2 = 0$
(a small boost and / or rotation).  Then the sum and integrals can be performed and the behavior for small and large $\mu r$ can be extracted.\footnote{For a massless scalar on the brane we must keep the
bulk scalar mass $\mu$ non-zero to avoid an IR divergence.}  This leads to
\be
b^2 \approx 0,\, m \approx 0 \, : \quad I_2 \approx \left\lbrace\begin{array}{ll}
\displaystyle {1 \over 6 \pi^2 \mu^2 r^2} & \quad \hbox{\rm as $\mu r \rightarrow 0$} \\[15pt]
\displaystyle {2 \over 3} \, e^{-2 \pi \mu r} & \quad \hbox{\rm as $\mu r \rightarrow \infty$}
\end{array}\right.
\ee

\section{Electron self-energy from a bulk graviton\label{sect:electron-graviton}}
\begin{figure}
\begin{center}
\includegraphics[height=3cm]{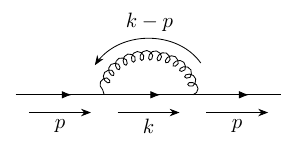}
\end{center}
\caption{Electron self-energy due to a bulk graviton loop.\label{fig:graviton}}
\end{figure}

The graviton is the most likely candidate for a bulk field.  It also provides an interesting contrast to the bulk scalars we have considered so far,
since it carries spin and has a non-renormalizeable coupling to the stress tensor on the brane.  For these reasons we consider corrections to
the electron self-energy induced by a bulk graviton loop.  The diagram is shown
in Fig.\ \ref{fig:graviton}.

Bulk gravitons in the large extra dimension scenario \cite{Antoniadis:1990ew,Arkani-Hamed:1998jmv,Antoniadis:1998ig}
have been considered in \cite{Giudice:1998ck} and we borrow several of their expressions.  We expand the 5-D metric
about flat space,
\be
g_{AB} = \eta_{AB} + {2 \over \overline{M}_5^{\,3/2}} h_{AB}
\ee
where $\overline{M}_5$ is the 5-D reduced Planck mass.  The brane-to-brane graviton propagator is
\be
\label{GravitonPropagator}
\raisebox{-0.75cm}{\includegraphics[height=1.5cm]{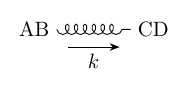}} \qquad\quad
{1 \over 2 \pi r} \sum_{n = -\infty}^\infty {i \over 2} \, {\eta_{AC} \eta_{BD} + \eta_{AD} \eta_{BC} - {2 \over 3} \eta_{AB} \eta_{CD}
\over k^2 - (k \cdot b + {n \over r})^2 - \mu^2 + i\epsilon}
\ee
where $k$ is the 4-D momentum, $n$ is the Kaluza-Klein momentum and we've introduced $\mu$ as an infrared regulator.
The propagator is written in de Donder gauge, $\xi = 1$ in the notation of \cite{Giudice:1998ck}.  We assume the graviton couples
to the 4-D stress tensor on the brane,
\bea
&& {\cal L} = \bar{\psi} \left(i \gamma^\mu \partial_\mu - m\right)\psi - {1 \over \overline{M}_5^{\, 3/2}} T^{\mu\nu} h_{\mu\nu} \big\vert_{z = 0} \\
\nonumber
&& T_{\mu\nu} = {i \over 4} \bar{\psi} \left(\gamma_\mu \partial_\nu + \gamma_\nu \partial_\mu\right)\psi
- {i \over 4} \left(\partial_\mu \bar{\psi} \gamma_\nu + \partial_\nu \bar{\psi} \gamma_\mu\right)\psi
\eea
which leads to the vertex
\be
\raisebox{-1.2cm}{\includegraphics[height=2.6cm]{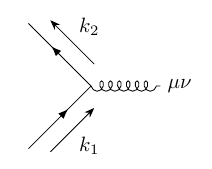}} \qquad
- {i \over 4 \overline{M}_5^{\,3/2}} \left[(k_1 + k_2)_\mu \gamma_\nu + (k_1 + k_2)_\nu \gamma_\mu\right]
\ee

So far the motion of the brane has only entered in the graviton propagator (\ref{GravitonPropagator}), in a manner exactly analogous to the
scalar propagator (\ref{MomentumSum}).  However the motion of the brane also enters in the effective 4-D coupling.  The Newtonian potential on a moving brane was studied
by Greene et al.\ in \cite{Greene:2011fm}, who found that the relation between the 4-D and 5-D
reduced Planck masses becomes\footnote{In \cite{Greene:2011fm} this was expressed in terms of Newton's constant,
$G_4 = G_5 / \gamma 2 \pi R$ with $G_4 = 1 / 8 \pi \overline{M}_4^{\,2}$ and $G_5 = 1 / 8 \pi \overline{M}_5^{\,3}$.}
\be
\overline{M}_4 = (\gamma 2 \pi R)^{1/2} \, \overline{M}_5^{\,3/2}
\ee
For a moving brane $r = \gamma R$ so the reduced 4-D Planck mass is
\be
\overline{M}_4 = (2 \pi r)^{1/2} \, \overline{M}_5^{\,3/2} = 2.4 \times 10^{18} \, {\rm GeV}
\ee
We take this relation to hold in general, i.e.\ even for a brane that is tilt-like rather than boost-like.

After all these preliminaries we are ready to evaluate the diagram in Fig.\ \ref{fig:graviton}.\footnote{There is another self-energy
diagram at one loop \raisebox{-0.4cm}{\includegraphics[height=1cm]{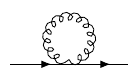}} but it does not induce Lorentz violation on the brane.}  The basic diagram is straightforward
to write down.
\be
i \Sigma = {1 \over 16 \pi r \overline{M}_5^{\,3}} \sum_{n = - \infty}^\infty \int {d^4k \over (2 \pi)^4} \,
{(k + p)^2 \gamma_\mu (\slashed{k} + m) \gamma^\mu + {1 \over 3} (\slashed{k} + \slashed{p}) (\slashed{k} + m) (\slashed{k} + \slashed{p}) \over
(k^2 - m^2 + i \epsilon) \big( (k-p)^2 - \big( (k - p) \cdot b + {n \over r}\big)^2 - \mu^2 + i \epsilon \big)}
\ee
With some Dirac algebra the numerator can be simplified so it is at most linear in Dirac matrices.  We Wick rotate, introduce Schwinger parameters, and perform the Gaussian integral over $k_E$.
It is convenient to do this in a frame in which the Euclidean vectors have components
\bea
\nonumber
&& b_E = (b_1,0,0,0) \\
&& p_E = (p_1, p_2,0,0) \\
\nonumber
&& k_E = (k_1,k_2,k_3,k_4)
\eea
Then we continue back to Lorentzian signature, restore Lorentz covariance, and expand in powers of $p$.
At zeroth order all terms are either Lorentz invariant or vanish because they are odd under $n \rightarrow -n$.  At first order in $p$, after switching from
a momentum sum to a winding sum, we find that many terms are either Lorentz invariant or vanish because they are odd under $w \rightarrow -w$.  Discarding all such terms we are left with a
Lorentz-violating contribution to the effective Lagrangian, ${\cal L} = i c_{\mu\nu} \bar{\psi} \gamma^\mu \partial^\nu \psi$ where\footnote{We rescaled the Schwinger parameters,
$s \rightarrow (\pi r w)^2 s$ and $t \rightarrow (\pi r w)^2 t$.}
\bea
\label{gravityc}
& & c_{\mu\nu} = - {1 \over 16 \pi^2} \, {1 \over (\pi r \overline{M}_4)^2} \left(b_\mu b_\nu - {1 \over 4} \eta_{\mu\nu} b^2\right) I_{\rm gravity} \\[5pt]
\nonumber
& & I_{\rm gravity} = {1 \over 6 \sqrt{\pi}} \sum_{w = 1}^\infty \int_0^\infty ds \int_0^\infty dt \, {9s^2 - 2st - 11t^2 + 5 s b^2 \over w^3 t^{1/2} (s+t)^6} \\
\nonumber
& & \hspace{2cm} \exp \left\lbrace -s(\pi m r w)^2 - t (\pi \mu r w)^2 - {s + t (1 - b^2) \over t(s+t)}\right\rbrace
\eea
We have written (\ref{gravityc}) as a product of
\begin{itemize}
\item
a loop factor ${1 \over 16 \pi^2}$,
\item
a dimensionless coupling ${1 \over (\pi r \overline{M}_4)^2}$
built from the effective radius $r$ and the reduced 4-D Planck mass $\overline{M}_4$,
\item
a symmetric traceless tensor structure $b_\mu b_\nu - {1 \over 4} \eta_{\mu\nu} b^2$,
\item
a function $I_{\rm gravity}$ of the dimensionless parameters $b^2$, $\pi m r$, $\pi \mu r$.
\end{itemize}

\begin{figure}
\begin{center}
\includegraphics[height=9cm]{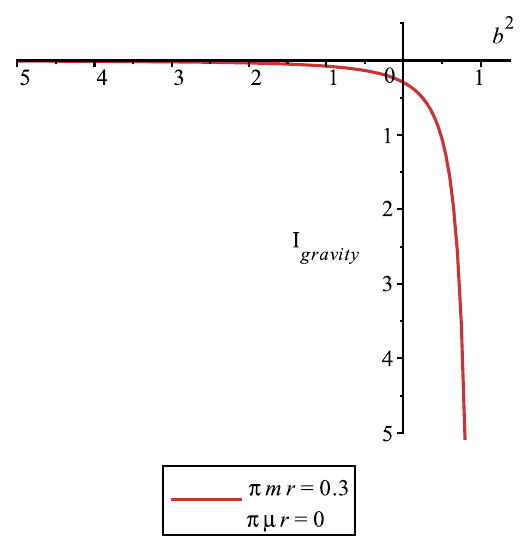}
\end{center}
\caption{The quantity $I_{\rm gravity}$ appearing in the electron self-energy due to a bulk graviton loop.  The function decreases
rapidly but has a finite limit as $b^2 \rightarrow 1$.\label{fig:Igravity}}
\end{figure}

The function $I_{\rm gravity}$ is shown in Fig.\ \ref{fig:Igravity}.  It simplifies if we set $m = 0$ (a massless fermion on the brane) and
$b^2 = 0$ (a small boost and / or rotation).  Then the sum and integrals can be performed and the behavior for large and small $\mu$
can be extracted.  For graviton loops there is no IR divergence, even for a massless fermion on the brane, and we find
\be
b^2 \approx 0,\, m \approx 0 \, : \quad I_{\rm gravity} \approx \left\lbrace\begin{array}{ll}
\displaystyle - {1 \over 4} \zeta(3) & \quad \hbox{\rm as $\mu r \rightarrow 0$} \\[12pt]
\displaystyle - {1 \over 3} \, (\pi \mu r)^2 e^{-2 \pi \mu r} & \quad \hbox{\rm as $\mu r \rightarrow \infty$}
\end{array}\right.
\ee

\section{Scalar self-energy from a bulk graviton\label{sect:scalar-graviton}}
Finally we consider corrections to the self-energy of a minimally-coupled scalar field due to a bulk graviton loop.  We assume the graviton couples
to the 4-D stress tensor on the brane,
\bea
&& {\cal L} = {1 \over 2} \partial_\mu \phi \partial^\mu \phi - {1 \over 2} m^2 \phi^2 - {1 \over \overline{M}_5^{\, 3/2}} T^{\mu\nu} h_{\mu\nu} \big\vert_{z = 0} \\
\nonumber
&& T_{\mu\nu} = \partial_\mu \phi \partial_\nu \phi - {1 \over 2} \eta_{\mu\nu} \big(\partial_\lambda \phi \partial^\lambda \phi - m^2 \phi^2\big)
\eea
which leads to the scalar -- graviton vertex
\be
\raisebox{-1.2cm}{\includegraphics[height=2.6cm]{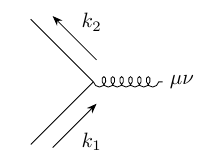}} \qquad
- {i \over \overline{M}_5^{\,3/2}} \left[k_{1 \mu} k_{2 \nu} + k_{1 \nu} k_{2 \mu} - \eta_{\mu\nu} (k_1 \cdot k_2 - m^2)\right]
\ee

\begin{figure}
\begin{center}
\includegraphics[height=3cm]{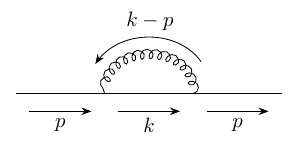}
\end{center}
\caption{Scalar self-energy due to a bulk graviton loop.\label{fig:scalar-graviton}}
\end{figure}

The diagram we wish to evaluate is shown in Fig.\ \ref{fig:scalar-graviton}.\footnote{There is another self-energy
diagram at one loop \raisebox{-0.3cm}{\includegraphics[height=1cm]{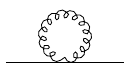}} but it does not induce Lorentz violation on the brane.}
With the brane-to-brane graviton propagator (\ref{GravitonPropagator}) the basic diagram is straightforward to write down.
\be
i \Sigma = {1 \over 4 \pi r \overline{M}_5^{\,3}} \sum_{n = - \infty}^\infty \int {d^4k \over (2 \pi)^4} \,
{4 k^2 p^2 + {4 \over 3} (k \cdot p)^2 + {8 \over 3} m^2 k \cdot p - {8 \over 3} m^4 \over
(k^2 - m^2 + i \epsilon) \big( (k-p)^2 - \big( (k - p) \cdot b + {n \over r}\big)^2 - \mu^2 + i \epsilon \big)}
\ee
Compared to the electron self-energy considered in section \ref{sect:electron-graviton} the main difference is in the contractions of the stress tensors in the numerator.  As is by now familiar we Wick rotate,
 introduce Schwinger parameters, and perform the Gaussian integral over $k_E$.  It is convenient to do this in a frame in which the Euclidean vectors have components
\bea
\nonumber
&& b_E = (b_1,0,0,0) \\
&& p_E = (p_1, p_2,0,0) \\
\nonumber
&& k_E = (k_1,k_2,k_3,k_4)
\eea
Then we continue back to Lorentzian signature, restore Lorentz covariance, switch from a momentum sum to a winding sum, and expand in powers of $p$.
At zeroth order in $p$ the expression is Lorentz invariant.  At first order in $p$ the result vanishes because all terms are odd under $w \rightarrow -w$.
At second order in $p$ many of the terms are Lorentz invariant.
Discarding all Lorentz-invariant terms we are left with a Lorentz-violating contribution to the effective Lagrangian, ${\cal L} = {1 \over 2} k_{\mu\nu} \partial^\mu \phi \partial^\nu \phi$
where\footnote{We rescaled the Schwinger parameters, $s \rightarrow (\pi r w)^2 s$ and $t \rightarrow (\pi r w)^2 t$.}
\bea
\label{gravityk}
& & k_{\mu\nu} = - {1 \over 16 \pi^2} \, {1 \over (\pi r \overline{M}_4)^2} \left(b_\mu b_\nu - {1 \over 4} \eta_{\mu\nu} b^2\right) I_{\rm gravity}^{\,{\rm scalar}} \\[5pt]
\nonumber
& & I_{\rm gravity}^{\,{\rm scalar}} = {4 \over 3 \sqrt{\pi}} \sum_{w = 1}^\infty \int_0^\infty ds \int_0^\infty dt \, {1 + 4 s (\pi m r w)^2 - 4 s^2 (\pi m r w)^4 \over w^3 t^{1/2} (s+t)^4} \\
\nonumber
& & \hspace{2cm} \exp \left\lbrace -s(\pi m r w)^2 - t (\pi \mu r w)^2 - {s + t (1 - b^2) \over t(s+t)}\right\rbrace
\eea
We have written (\ref{gravityk}) as a product of
\begin{itemize}
\item
a loop factor ${1 \over 16 \pi^2}$,
\item
a dimensionless coupling ${1 \over (\pi r \overline{M}_4)^2}$
built from the effective radius $r$ and the reduced 4-D Planck mass $\overline{M}_4$,
\item
a symmetric traceless tensor structure $b_\mu b_\nu - {1 \over 4} \eta_{\mu\nu} b^2$,
\item
a function $I_{\rm gravity}^{\,{\rm scalar}}$ of the dimensionless parameters $b^2$, $\pi m r$, $\pi \mu r$.
\end{itemize}

\begin{figure}
\begin{center}
\includegraphics[height=9cm]{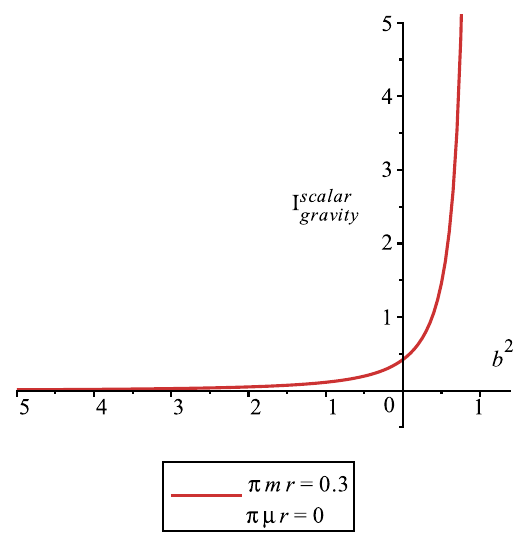}
\end{center}
\caption{The quantity $I_{\rm gravity}^{\,{\rm scalar}}$ appearing in the scalar self-energy due to a bulk graviton loop.  The function increases
rapidly but has a finite limit as $b^2 \rightarrow 1$.\label{fig:Iscalar-gravity}}
\end{figure}

The function $I_{\rm gravity}^{\,{\rm scalar}}$ is shown in Fig.\ \ref{fig:Iscalar-gravity}.  It simplifies if we set $m = 0$ (a massless scalar on the brane) and
$b^2 = 0$ (a small boost and / or rotation).  Then the sum and integrals can be performed and the behavior for large and small $\mu$
can be extracted.  There is no IR divergence in $I_{\rm gravity}^{\,{\rm scalar}}$, even for a massless scalar on the brane, and we find
\be
b^2 \approx 0,\, m \approx 0 \, : \quad I_{\rm gravity}^{\,{\rm scalar}} \approx \left\lbrace\begin{array}{ll}
\displaystyle {1 \over 3} \zeta(3) & \quad \hbox{\rm as $\mu r \rightarrow 0$} \\[12pt]
\displaystyle {4 \over 9} \, (\pi \mu r)^2 e^{-2 \pi \mu r} & \quad \hbox{\rm as $\mu r \rightarrow \infty$}
\end{array}\right.
\ee

\section{Conclusions\label{sect:conclusions}}
In this work we considered a braneworld which is moving or spiraling around a compact extra dimension which we take to be a
circle of radius $R$.  The configuration
is described by an effective radius $r$ for the compactification and a 4-vector $b^\mu$ that spontaneously breaks the Lorentz symmetry of the brane worldvolume.
\bea
\nonumber
&& r = \left\lbrace
\begin{array}{ll}
\gamma R & \quad \hbox{\rm boost-like} \\[2pt]
R & \quad \hbox{\rm null} \\[2pt]
\cos \theta R & \quad \hbox{\rm tilt-like}
\end{array}
\right. \\[5pt]
&& b^\mu = \left\lbrace
\begin{array}{ll}
(-\beta,0,0,0) & \quad \hbox{\rm boost-like} \\[2pt]
(-\lambda,\lambda,0,0) & \quad \hbox{\rm null} \\[2pt]
(0,\tan \theta, 0,0) & \quad \hbox{\rm tilt-like}
\end{array}\right.
\eea
Loops of bulk fields are sensitive to the parameter $b^\mu$ and can induce Lorentz-violating terms in the 4-D effective
action.  We explored this, emphasizing the dimension-4 terms which correct the electron self-energy and the electron -- photon
vertex.
\be
{\cal L} = i c_{\mu\nu} \bar{\psi} \gamma^\mu D^\nu \psi \qquad\quad D_\mu = \partial_\mu - i e A_\mu
\ee
The one-loop coefficients $c_{\mu\nu} \sim b_\mu b_\nu - {1 \over 4} \eta_{\mu\nu} b^2$ due to bulk scalars and gravitons
are given in (\ref{c}), (\ref{gravityc}).  There are stringent experimental bounds on Lorentz violation, reviewed in \cite{Kostelecky:2008ts}.  For the electron, for example,
laboratory bounds on the dimensionless coefficients $c_{\mu\nu}$ have reached the level of $\sim 10^{-21}$ \cite{Dreissen:2022tks}.

The Standard Model Extension is a general framework for incorporating Lorentz violation and provides many effects to explore.  In addition to the QED effects mentioned above,
we considered Lorentz-violating corrections to the self-energy of a scalar field, ${\cal L} = {1 \over 2} k_{\mu\nu} \partial^\mu \phi \partial^\nu \phi$
with coefficients $k_{\mu\mu}$ given in (\ref{k}) and (\ref{gravityk}).  Taking the scalar field as a proxy for the Higgs field the experimental bounds on $k_{\mu\nu}$ are
surprisingly good \cite{Kostelecky:2008ts}, having reached the level of $10^{-12}$ -- $10^{-20}$ \cite{Hernandez-Juarez:2018dkx} or $10^{-13}$ -- $10^{-27}$ \cite{Anderson:2004qi}.

While many similar calculations could be done, there are also theoretical issues worth exploring.  In particular it would be interesting to understand soft emission from a moving
braneworld.  This should be related to the infrared behavior of the diagrams we have considered.  For example for $\mu^2 = 0$ the vertex correction (\ref{vertex_diagram}) has an
IR divergence when $p_1^2 = p_2^2 = m^2$ that should cancel against soft emission in suitable inclusive observables.

Any signal for Lorentz violation in the present epoch would be of the utmost significance.  One can also entertain the idea that, although Lorentz-violating effects
are extremely small today, they may have been larger in the early universe.  Perhaps a
braneworld was highly boosted in the early universe and only slowed and stabilized with time.  Could the attendant
violation of Lorentz symmetry in the early universe leave an observable imprint on cosmology?

\bigskip
\goodbreak
\centerline{\bf Acknowledgements}
\noindent
We are grateful to Brian Greene, Janna Levin, Alexios Polychronakos and Massimo Porrati for valuable discussions.  DK is supported by U.S.\ National
Science Foundation grant PHY-2112548.

\appendix
\section{More on lightlike $b^\mu$\label{appendix:null}}

\begin{figure}
\begin{center}
\includegraphics[height=4cm]{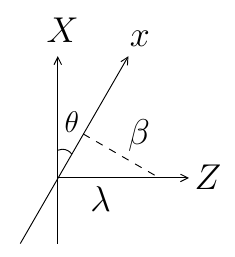}
\end{center}
\caption{The case of light-like $b^\mu$.  The angle between the brane and the $X$ axis is $\theta = \tan^{-1} \lambda$.  The brane moves along the $Z$ axis with velocity $\lambda$; the perpendicular component of the velocity is denoted $\beta$.\label{fig:null}}
\end{figure}

Since the geometry of the null case may be a little unfamiliar we give some further explanation.  According to (\ref{J-Z2}) a brane at $z = 0$ has $Z = \lambda X^+$, so the brane spirals around the $S^1$ as one
moves in the $X^+$ direction.  At $T = 0$ the brane is located at $Z = \lambda X$, which means it has been rotated in the $XZ$ plane by an angle $\theta = \tan^{-1} \lambda$.
Setting $X = 0$ we have $Z = \lambda T$, which means the brane is moving along the $Z$ axis with velocity $\lambda$.  (As in the ``closing scissors'' effect this velocity can be arbitrarily large.)  However it's the component of the velocity perpendicular to the brane which is physically relevant, and as can be seen in Fig.\ \ref{fig:null} this is given by
\be
\beta = \lambda \cos \theta = {\lambda \over \sqrt{1 + \lambda^2}} \qquad \gamma = \sqrt{1 + \lambda^2}
\ee
Thus we can summarize the light-like case as a combination of a boost and a rotation with\footnote{We're grateful to Massimo Porrati for pointing out this relation.}
\be
\gamma \beta = \tan \theta = \lambda
\ee

\begin{figure}
\begin{center}
\includegraphics[height=4cm]{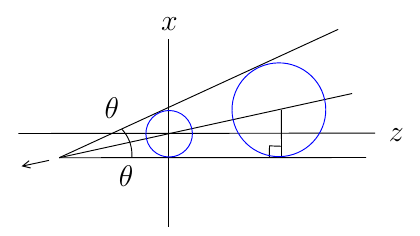}
\end{center}
\caption{At fixed $t$ the future lightcones of the image charges form circles in the $xz$ plane.  Their envelope forms a cone which moves in the direction indicated by the arrow.  The lower part of the envelope is parallel to
the $z$ axis and moves downward at the speed of light.  The upper part of the envelope intersects the $x$ axis at $x = {1 + \lambda^2 \over 1 - \lambda^2} \, t$.\label{fig:xz}}
\end{figure}

We'd like to understand what a causal signal in the bulk looks like on such a brane.  Following the analysis in \cite{Greene:2022uyf} we consider a bulk signal sent out from the origin $T = X = {\bf Y} = Z = 0$ and ask where its
future lightcone intersects the brane.  It's simplest to work in the covering space where the origin corresponds to an infinite series of image charges located along the $Z$ axis at
\be
Z_w = 2 \pi R w \qquad w \in {\mathbb Z}
\ee
In the bulk the future lightcones of the image charges are given by
\be
X^+ X^- - \vert {\bf Y} \vert^2 - (Z - 2 \pi R w)^2 = 0
\ee
Using (\ref{J-Z2}) to switch to co-moving coordinates and recalling that for this case $r = R$, the future lightcones are given by
\be
(x - 2 \pi r w \lambda)^2 + \vert {\bf y} \vert^2 + (z - 2 \pi r w)^2 = (t + 2 \pi r w \lambda)^2
\ee
At fixed $t$ we see that the future lightcones are spheres of radius $t + 2 \pi r w \lambda$ centered at
\be
x = 2 \pi r w \lambda \qquad {\bf y} = 0 \qquad z = 2 \pi r w
\ee
The situation in the $xz$ plane is shown in Fig.\ \ref{fig:xz}.  The centers of the spheres lie along the line $x = \lambda z$.  Their envelope defines a cone with opening
angle $\theta = \tan^{-1} \lambda$.  The tip of the cone is located at\footnote{At time $t$ the tip, where the radius shrinks to zero, corresponds to a fictitious image charge with
$w = - t / 2 \pi r \lambda$.}
\be
x = -t \qquad z = - t / \lambda
\ee
The bottom part of the envelope is horizontal and intersects the $x$ axis at
\be
x = -t
\ee
This means bulk signals propagate in the $-x$ direction at the speed of light.  The top part of the envelope, on the other hand, intersects the $x$ axis at
\bea
\nonumber
x & = & -t + {t \over \lambda} \tan (2 \theta) \\[3pt]
& = & {1 + \lambda^2 \over 1 - \lambda^2} t
\eea
Thus bulk signals propagate in the $+x$ direction with speed ${1 + \lambda^2 \over 1 - \lambda^2}$ in agreement with (\ref{plusx}).  For $0 < \lambda < 1$ there is superluminal propagation in
the $+x$ direction.  At $\lambda = 1$ the velocity diverges and propagation is instantaneous.  For $\lambda > 1$ the velocity is negative, which can be thought of as a signal from the origin that is traveling in the $+x$
direction but backwards in time.  Alternatively it can be thought of as a signal going forward in time that was emitted in the far past at $x = + \infty$, destined to reach the origin at $t = 0$.
This can be seen geometrically in Fig.\ \ref{fig:xz} from the fact that the range $1 < \lambda < \infty$ corresponds to $\pi / 2 < 2 \theta < \pi$.

If we include the transverse directions, then at time
$t$ and position $x = z = 0$ the envelope extends into the transverse directions a distance
\bea
\nonumber
\vert {\bf y} \vert & = & \sqrt{t^2 + \left({t \over \lambda}\right)^2} \, \tan \theta \\
& = & \sqrt{1 + \lambda^2} \, t
\eea
Thus bulk signals propagate in the transverse directions at a superluminal speed $\sqrt{1 + \lambda^2}$.  To relate this to (\ref{vperp}), note that a particle
moving in the transverse directions has $x = 0$, which means $x^+ = x^- = t$ and hence $v^- = 1$.  Then (\ref{vminus}) becomes $\vert {\bf v}_\perp \vert^2 + {(n/r)^2 + \mu^2 \over (k^+)^2} = 1 + \lambda^2$, so the transverse velocity is bounded above by $\sqrt{1 + \lambda^2}$.

\providecommand{\href}[2]{#2}\begingroup\raggedright\endgroup

\end{document}